\documentclass{elsart}
\usepackage{graphicx}
\usepackage{dcolumn}
\usepackage{bm}
\usepackage{amsmath}
\usepackage{amsfonts}
\usepackage{amssymb}
\begin{document}
\begin{frontmatter}
\title{Quantum search with resonances}
\author{A. Romanelli, }
\author{A. Auyuanet\thanksref{PEPE}},
\author{and R. Donangelo\thanksref{UFRJ}}
\address{Instituto de F\'{\i}sica, Facultad de Ingenier\'{\i}a\\
Universidad de la Rep\'ublica\\ C.C. 30, C.P. 11000, Montevideo, Uruguay}
\thanks[PEPE]{Corresponding author. \textit{E-mail address:}
auyuanet@fing.edu.uy}
\thanks[UFRJ]{Permanent address: Instituto de F\'{\i}sica, Universidade Federal do Rio de
Janeiro\\
C.P. 68528, 21941-972 Rio de Janeiro, Brazil}
\date{\today}
\begin{abstract}
\vspace
{0.2cm} \\
We present a continuous time quantum search algorithm analogous to Grover's. 
In particular, the optimal search time for this algorithm is proportional to $\sqrt{N}$,
where $N$ is the database size. 
This search algorithm can be implemented using any Hamiltonian with a discrete energy 
spectrum through excitation of resonances between an initial and the searched state.
This algorithm is robust and, as in the case of Grover's, it allows for an error 
$O(1/\sqrt{N})$ in the determination of the searched state. 
A discrete time version of this continuous time search algorithm is built, and the 
connection between the search algorithms with discrete and continuous times is established.
\end{abstract}
\end{frontmatter}

\section{Introduction}

In the last twenty years the attention of researchers of several different
areas has been attracted towards quantum computation~\cite{Feynman,Chuang}.
This field of knowledge presents new scientific challenges both from the
theoretical and the experimental points of view. Part of the theoretical
challenge is to learn how to work with quantum properties to obtain new and
more efficient algorithms.

A quantum algorithm can be seen as a definite sequence of unitary transformations 
acting over a quantum state, in some Hilbert space. Its size
is proportional to the number of elementary unitary transformations of which
the algorithm is composed. Concepts like interference phenomena, quantum
measurements, resonances, quantum parallelism, amplification techniques, etc.,
are employed by this new computation science. However, relatively few quantum
algorithms were created; among them, Shor's and Grover's~\cite{Shor,Grover}
algorithms are the best known. Shor's algorithm decomposes a number in its
prime factors more efficiently than any known classical algorithm. To achieve
this it uses quantum parallelism, quantum Fourier transforms, and the
properties of quantum measurements. Grover's search algorithm locates a marked
item in an unsorted list of $N$ elements in a number of steps proportional to
$\sqrt{N}$, instead of $O(N)$ as in the classical case. It performs a unitary
transformation of the quantum state which increases the likelihood that the
marked state of interest will be measured at the output (amplification
technique). It has been proven that there are neither quantum nor classical
algorithms that can perform faster such an unstructured search~\cite{Boyer}.

In this work we present a continuous time quantum search algorithm, which is
controlled by a time dependent Hamiltonian. In particular there are not
unitary operators that are applied at certain time steps, as in Grover's
algorithm. Unlike other authors~\cite{Farhi,Childs} who studied continuous
time search algorithms, the most relevant characteristic of our model is the
use of quantum resonances, showing explicitly Grover's assertion that his
algorithm is a resonance phenomenon~\cite{Grover2}. We should remark that
a suggestion to employ resonances in quantum computing was made by M. Zak in~\cite{Zak}. 
The paper is organized as follows. In the next section we develop the continuous 
time search model. In section 3 we present examples of use of this model for two 
typical Hamiltonians. Then we show, in section 4, the connection of this continuous 
time with a discrete time search algorithm. 
In the last section, we make some concluding remarks.

\section{Resonances}

\label{sec:resonance}

The evolution of the wavefunction $|\Psi(t)\rangle$ satisfies the
Schr\"{o}dinger equation,%

\begin{equation}
i\frac{\partial|\Psi(t)\rangle}{\partial t}=H\text{ }|\Psi(t)\rangle,
\label{Schrodinger}%
\end{equation}

where $H=H_{0}+V(t)$ and we have taken Planck's constant $\hbar=1$. Here
$H_{0}$ is a known nondegenerate time-independent Hamiltonian with a discrete
energy spectrum, and $V(t)$ is a time dependent potential that will be
defined below. We should point out that the extension to the degenerate case
is quite straightforward, as in the case of Grover's algorithm for several
equally marked items.

Let us consider the normalized eigenstates $\left\{ |n\rangle\right\} $ and
eigenvalues $\left\{  \varepsilon_{n}\right\} $ for the unperturbed
Hamiltonian. These sets can be finite or infinite, depending on the nature of
$H_{0}$. We now consider a subset \textbf{N} of $\left\{ |n\rangle\right\}$ 
formed by $N$ elements on which we shall apply the search algorithm. 
We take a known eigenstate $|j\rangle$, with eigenvalue $\varepsilon_{j}$, 
as the initial state of the system. This initial state does not belong to 
the search set \textbf{N}.

Let us call $|s\rangle$ the unknown searched state whose energy 
$\varepsilon _{s}$ is known. This knowledge is equivalent to `mark' the 
searched state in Grover's algorithm. Our task is to find the eigenvector 
$|s\rangle$ which transition energy from a given initial state $|j\rangle$ 
is $\omega_{sj}\equiv\varepsilon_{j}-\varepsilon_{s}$. 
We propose the following potential $V$, which, as it may be easily verified, 
produces a resonance between the initial and the searched states,

\begin{equation}
V(t)=\left|  p\right\rangle \left\langle j\right|  \exp\left(  i\omega
_{sj}t\right)  +\left|  j\right\rangle \left\langle p\right|  \exp\left(
-i\omega_{sj}t\right)  ,
\label{potential}
\end{equation}

where $\left|  p\right\rangle \equiv\frac{1}{\sqrt{N}} 
{\displaystyle \sum\limits_{n\in{\textbf{N}}}}|n\rangle$ is an unitary vector, 
that can be interpreted as the average of the set of vectors \textbf{N}. 
This definition assures that the interaction potential $V$ is hermitian, that 
the transition probabilities
from state $|j\rangle$ to any state of the set \textbf{N} are all equal, 
$W_{nj}\equiv\left|  \left\langle n|V(t)|j\right\rangle \right| ^{2}=1/N$, 
and finally that the sum of the transition probabilities verifies
${\displaystyle\sum\limits_{n\in\mathbf{N}}}W_{nj}=1$. 

Let us express $|\Psi(t)\rangle$ as an expansion in the eigenstates
$\{|n\rangle\}$ of $H_{0}$,
$|\Psi(t)\rangle=\sum_{m}a_{m}(t)\exp\left( -i\varepsilon_{m}t\right) |m\rangle$,
where the expansion coefficients depend on time. Replacing the above expression
of $|\Psi(t)\rangle$ in eq.(\ref{Schrodinger}), we obtain the following set of
equations for the amplitudes $a_{m}(t)$

\begin{equation}
\frac{da_{n}(t)}{dt}=-i\sum_{m}\left\langle n|V(t)|m\right\rangle a_{m}%
(t)\exp\left(  -i\omega_{nm}t\right)  ,\label{dynamics}%
\end{equation}
where $\omega_{nm}=\varepsilon_{m}-\varepsilon_{n}$ are the Bohr frequencies.
Combining eqs.(\ref{potential}) and (\ref{dynamics}), we find,
\begin{equation}
\frac{da_{n}(t)}{dt}=0\ , \label{dynamics1}%
\end{equation}
if $n\notin\mathbf{N}$ and $n\neq j$; and%
\begin{align}
\frac{da_{n}(t)}{dt}  &  =-\frac{i}{\sqrt{N}}
\genfrac{\{}{.}{0pt}{}{{}}{{}}%
\left(  1-\delta_{nj}\right)  a_{j}(t)\exp\left[  +i\left(  \omega_{jn}%
+\omega_{sj}\right)  t\right] \nonumber\\
&  \makebox[2.5cm]{}\left.  +\delta_{nj}\sum_{m\in\mathbf{N}}a_{m}%
(t)\exp\left[  -i\left(  \omega_{jm}+\omega_{sj}\right)  t\right]  \right\}
\ ,\label{dynamics2}%
\end{align}
if $n\in\mathbf{N}$ or $n=j$.

Before solving numerically the set of eqs.(\ref{dynamics1}) and
(\ref{dynamics2}), it is important to understand the qualitative behavior of
these equations. Note that there are two time scales involved, a fast scale
associated to the Bohr frequencies, and a slow scale associated to the
amplitudes $a_{n}(t)$. Integrating the previous equations in a time interval
greater than the fast scale, the most important terms are those that have a
very small phase, as the others average to zero. In this approximation the
previous set of equations becomes
\begin{align}
\frac{da_{j}(t)}{dt}  &  \simeq-\frac{i}{\sqrt{N}}a_{s}(t)\ ,\\
\frac{da_{s}(t)}{dt}  &  =-\frac{i}{\sqrt{N}}a_{j}(t),\label{dynamics3}\\
\frac{da_{n}(t)}{dt}  &  \simeq0\text{ for all }n\neq s,j\ .
\end{align}
These equations represent two oscillators that are coupled so that their
population probabilities alternate in time. As we notice the coupling is
established between the initial and the searched for state. Solving those
equations with initial conditions $a_{j}(0)=1$, $a_{s}(0)=0$ we obtain
\begin{align}
P_{j}  &  \simeq\cos^{2}(\Omega\ t)\ ,\\
P_{s}  &  \simeq\sin^{2}(\Omega\ t)\ , \label{psearched}%
\end{align}
where $\Omega=\frac{1}{\sqrt{N}}$. It is important to notice that this
approach is valid only if all the Bohr frequencies verify $\omega_{nm}%
\gg\Omega$. If we let the system evolve during a time $\tau\equiv\frac{\pi}%
{2}\sqrt{N}$ , and we make a measurement immediately after that, the
probability to obtain the searched state is one. For the case that the
previous approximations are true, our method behaves qualitatively like Grover's,
but the time $\tau$ is twice the search time of Grover's algorithm \cite{Chuang}. 
This difference is not relevant since the resonance potential eq.(\ref{potential}) 
can be renormalized with a constant factor $V_{0}$ as long as it does not
invalidate the previous approximations, $\omega_{nm}\gg\Omega$, where now
$\Omega=V_{0}/\sqrt{N}$.

\section{Numerical results}
\label{sec:Numerical results}
We have integrated numerically eqs.(\ref{dynamics1}) and (\ref{dynamics2}) for
two Hamiltonians $H_{0}$, namely for the quantum harmonic oscillator and for
the quantum rotor in 2D.
The eigenvalues for these Hamiltonians are
$\varepsilon_{m}=\varepsilon _{0}(m+\frac{1}{2})$ and
$\varepsilon_{m}=\varepsilon_{0}m^{2}$ respectively.
In both cases we take $\varepsilon_{0}=1$ and initial conditions $a_{j}(0)=1$,
$a_{n}(0)=0$, for all $n\neq j$. The calculations were performed using a
standard fourth order Runge-Kutta algorithm. Choosing an arbitrary eigenvalue
for the energy of the searched state, we follow the dynamics of the set
\textbf{N}. We have verified, for several values of $N$ , that the most
important coupling is between the initial and the searched state, other
couplings being negligible, as discussed in the previous section.
\vspace{0.5cm} \newline \begin{figure}[h]
\begin{center}
\includegraphics[scale=0.6]{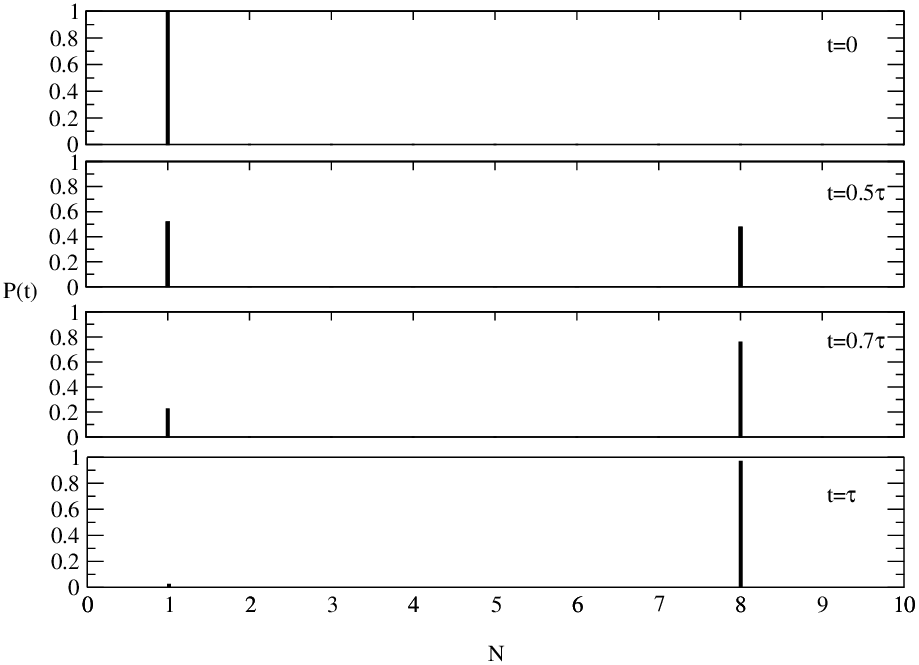}
\end{center}
\caption{Time evolution of the probability distribution of the initial and the
search set states, for the quantum rotor, for $\mathbf{N}=\left\lbrace 2,...
,10\right\rbrace $. The initial state was taken to be $j=1$, and the
searched state $s=8$. The full line shows the probability of the initial and
the searched state. For all the others states of the set the probabilities
take negligible values.}%
\label{evol}%
\end{figure}

\begin{figure}[ht]
\begin{center}
\includegraphics[scale=0.7]{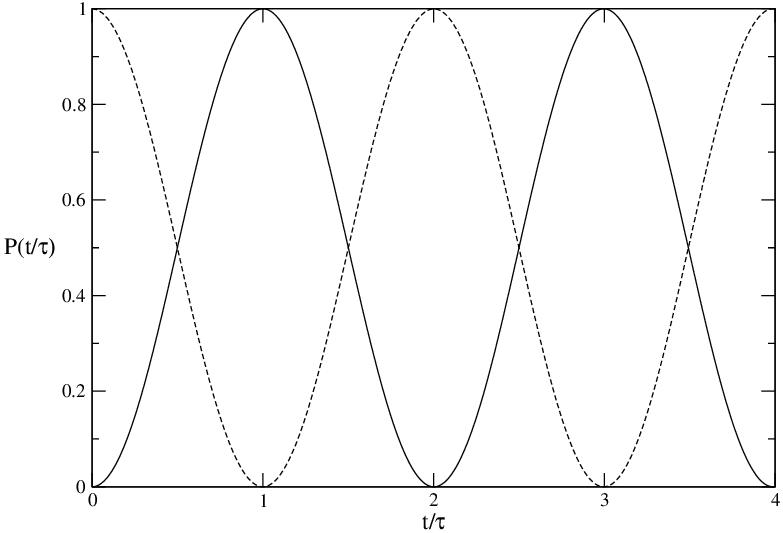}
\end{center}
\caption{ Probability distribution for the initial (dashed line) and the
searched (full line) states as a function of time for the quantum rotor
$H_{0}$ in the case of a data set of size $N=20$}%
\label{rotator}%
\end{figure}

\begin{figure}[htb]
\begin{center}
\includegraphics[scale=0.7]{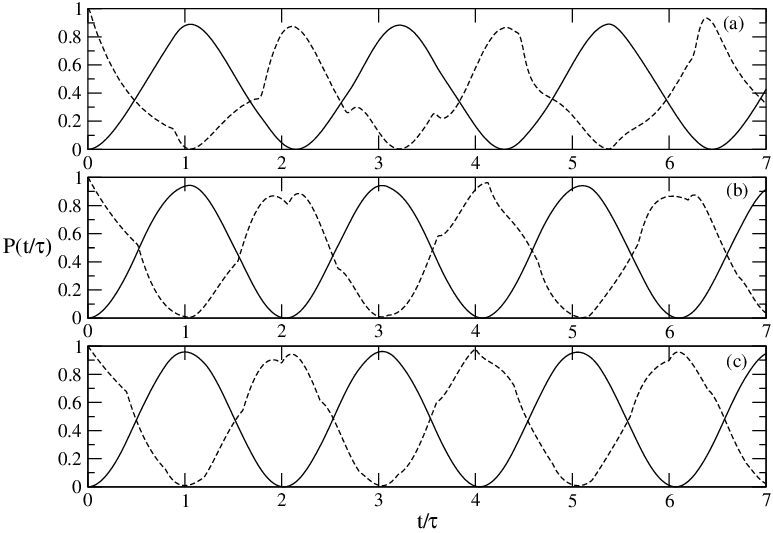}
\end{center}
\caption{Probability distribution for the harmonic oscillator $H_{0}$ as a
function of time. The dashed line corresponds to the initial state and the
full line to the searched state. The size of the searched set are: (a) $N=20$,
(b) $N=60$ , (c) $N=100.$
\vspace{1.5cm}}%
\label{osarm}%
\end{figure}

The probability distribution of the search set and the initially
loaded state are shown for the rotor case in Fig.\ref{evol}. Each panel
shows the probability distribution at different times. It is clear that the
total probability flows between the initial and the searched states. There is
a characteristic time at which the probability of the searched state is
maximum and its value is very near one. This time agrees with our theoretical
prediction, $\tau$. Fig.\ref{rotator} shows the oscillation of the probability
flux between the initial and the searched states as a function of time for the
quantum rotor. The temporal evolution was normalized for the characteristic
time $\tau$. Fig.\ref{osarm} shows a similar calculation for the harmonic
oscillator. Here we have considered three sizes for the data set. Note two
remarkable differences between Figs.\ref{rotator} and \ref{osarm}: for the
rotor the maximum values of the probabilities are one, while for the
harmonic oscillator they are less than one, and decrease with time.
Furthermore, while for the rotor the maxima are located in agreement with
the theoretical prediction, for the harmonic oscillator the maxima show an
increasing shift to the right of the predicted values as the search algorithm
evolves. 

\begin{figure}[htbt]
\begin{center}
\includegraphics[scale=0.32]{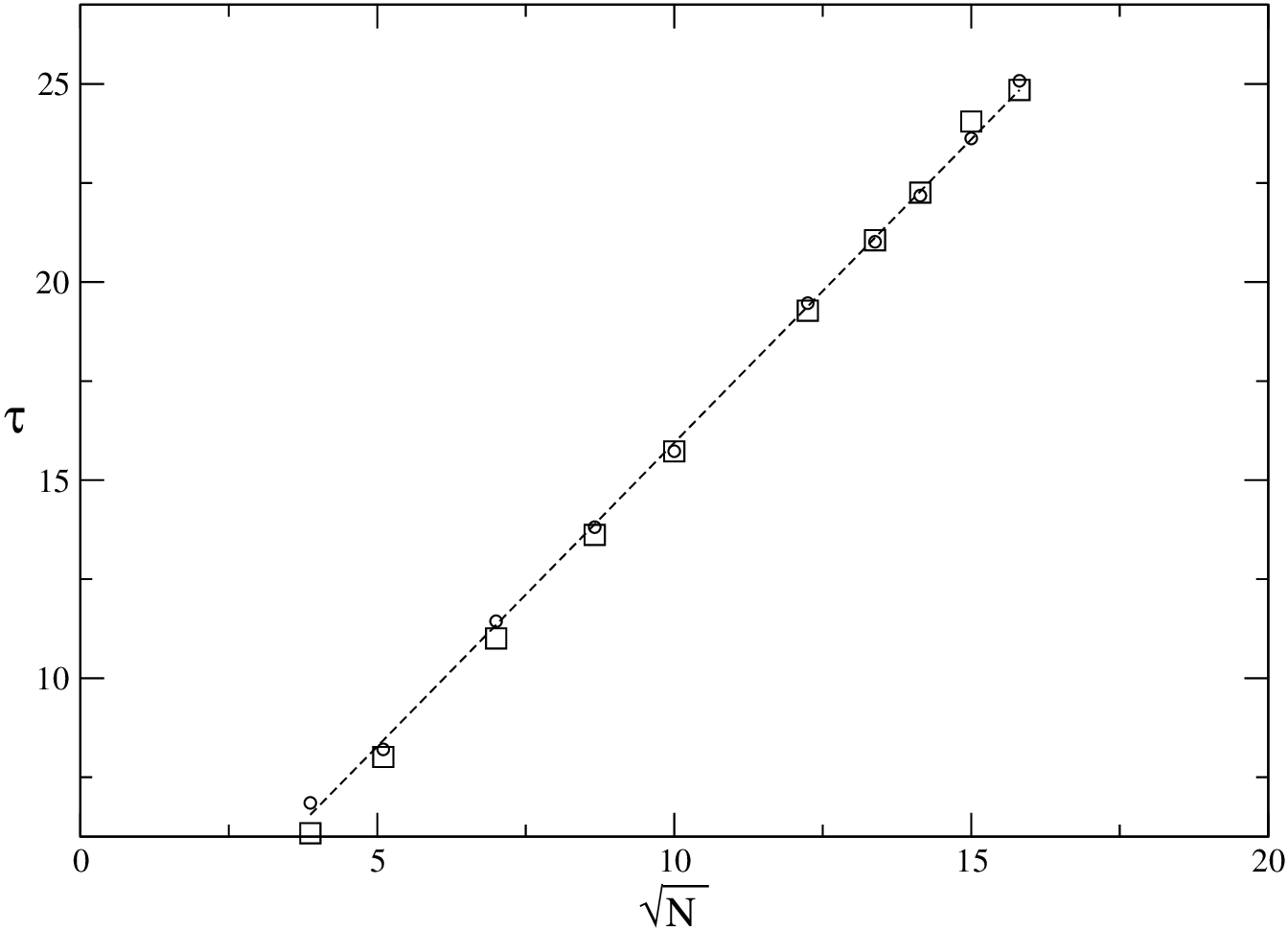}
\end{center}
\caption{Time $\tau$ at which the probability of the searched state is maximum
as a function of the square root of the dimension of the search set. Squares
correspond to the harmonic oscillator, and circles to the rotor.}%
\label{period}%
\end{figure}

The differences between the two Hamiltonians mentioned above are due to the 
properties of their respective energy spectra:
while the energy eigenvalues of the rotor increase quadratically with the
quantum number, for the harmonic oscillator they increase linearly.
Consequently, the condition $\Omega<<\omega_{nm}$, is better satisfied for the
spectrum of the rotor than for the harmonic oscillator. As a consequence the
agreement with the theoretical approximation is worse in this last case.
However, as $\Omega=\frac{1}{\sqrt{N}}$ , if the size of the data set is
increased, the agreement for the harmonic oscillator with the behavior
predicted by eq.(\ref{psearched}) is much improved, as shown in
Fig.\ref{osarm}. Increasing $N$ leads to a decrease in the relative distance
between the energy spectrum values of $H_{0}$, allowing for a better
resolution. In general, for a given Hamiltonian $H_{0}$ with a discrete energy
spectrum, the improvement obtained increasing $N$ becomes negligible after
some saturation value. Above this saturation value the algorithm presented
here, has the same characteristics as Grover's algorithm. \newline
\vspace{0.5cm} \newline \newline \vspace{1.5cm} 
\newline The proportionality between the characteristic time
$\tau$ and $\sqrt{N}$ for the rotor and the harmonic oscillator, is verified
in Fig.\ref{period}.
\newline 
\newline \\
Taking $N$ greater than its saturation value, one can test the robustness of this search
algorithm with respect to imprecisions in the eigenvalue of the searched state. 
Let us replace $\omega_{sj}$ by $\widetilde {\omega}_{sj}\equiv\omega_{sj}+\delta$ in
eqs.(\ref{dynamics2}), where we take $\delta$ smaller than $\Omega$ .

Within the this approximation the probability of the searched state evaluated
in $t=\tau$ is,
\begin{equation}
P_{s}(N,\delta)\ =\left[  \frac{\sin\left(  \frac{\pi}{2}\sqrt{1+\frac
{\delta^{2}N}{4}}\right)  }{\sqrt{1+\frac{\delta^{2}N}{4}}}\right]
^{2}.\label{senoc}%
\end{equation}
\newline We define the resonance width $\Delta$, as the value of $\delta$ for
which the probability of the searched state falls to half of the value for
$\delta=0$. \vspace{0.5cm}\newline \begin{figure}[h]
\begin{center}
\includegraphics[scale=0.33]{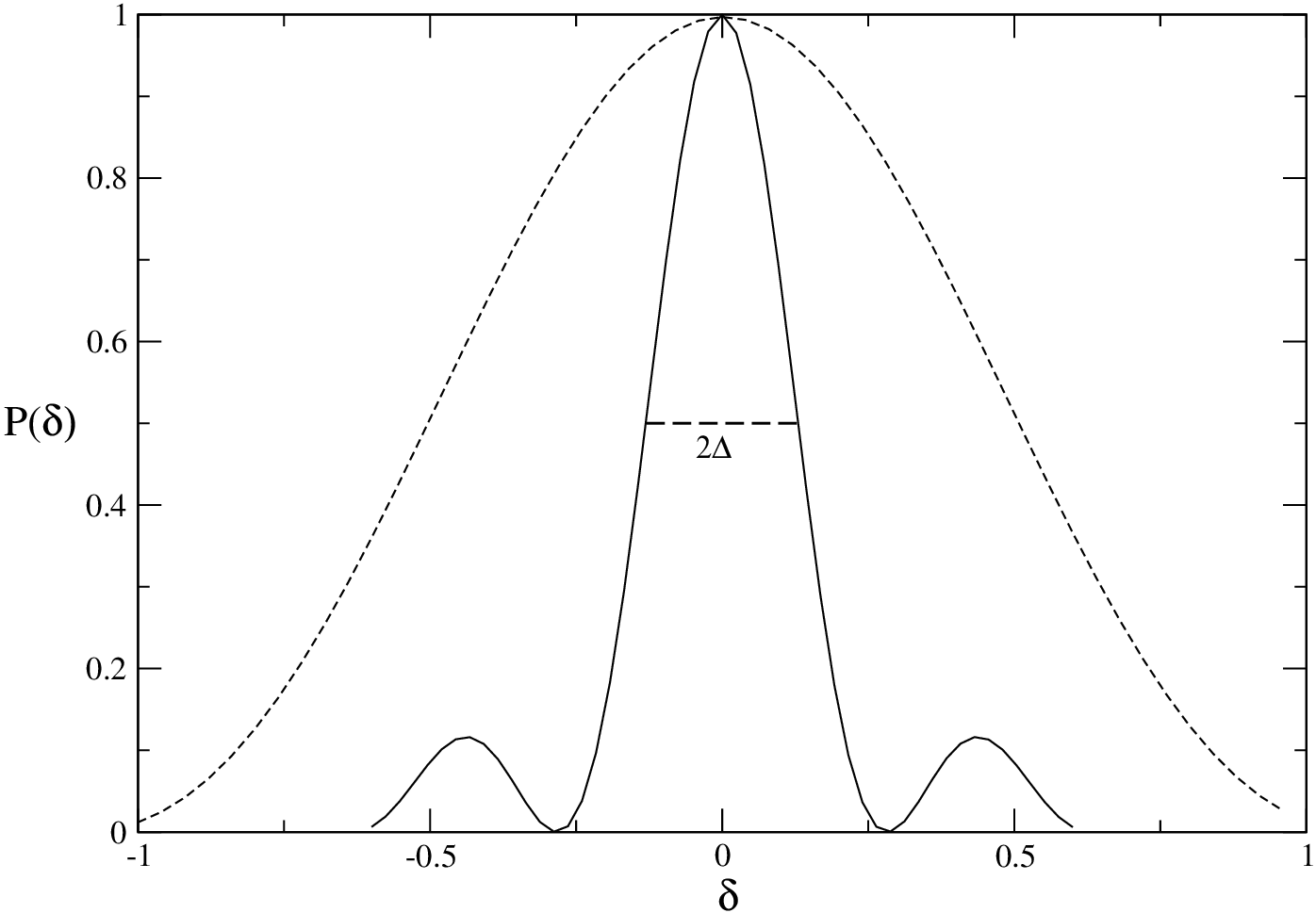}
\end{center}
\caption{Probability to find the searched state as a function of the energy
error $\delta$ , for the rotor , with $N=10$ (dashed line) and $N=100$ (full
line). $\Delta$ is the width of the resonance.}%
\label{width}
\end{figure} \\
\vspace{0.8cm}\\
The calculation of the probability of the searched state
as a function of $\delta$ is presented in Fig.\ref{width}. It can be observed
that the curves are in agreement with eq.(\ref{senoc}), showing a symmetrical
behavior about $\delta=0$ and a sharpness that depends on $N$. For small
$\delta$ the probability of the searched state remains large enough. Then, we
deduce that a small error in $\omega_{sj}$ does not affect drastically the
search algorithm.\\
\begin{figure}[h]
\begin{center}
\includegraphics[scale=0.35]{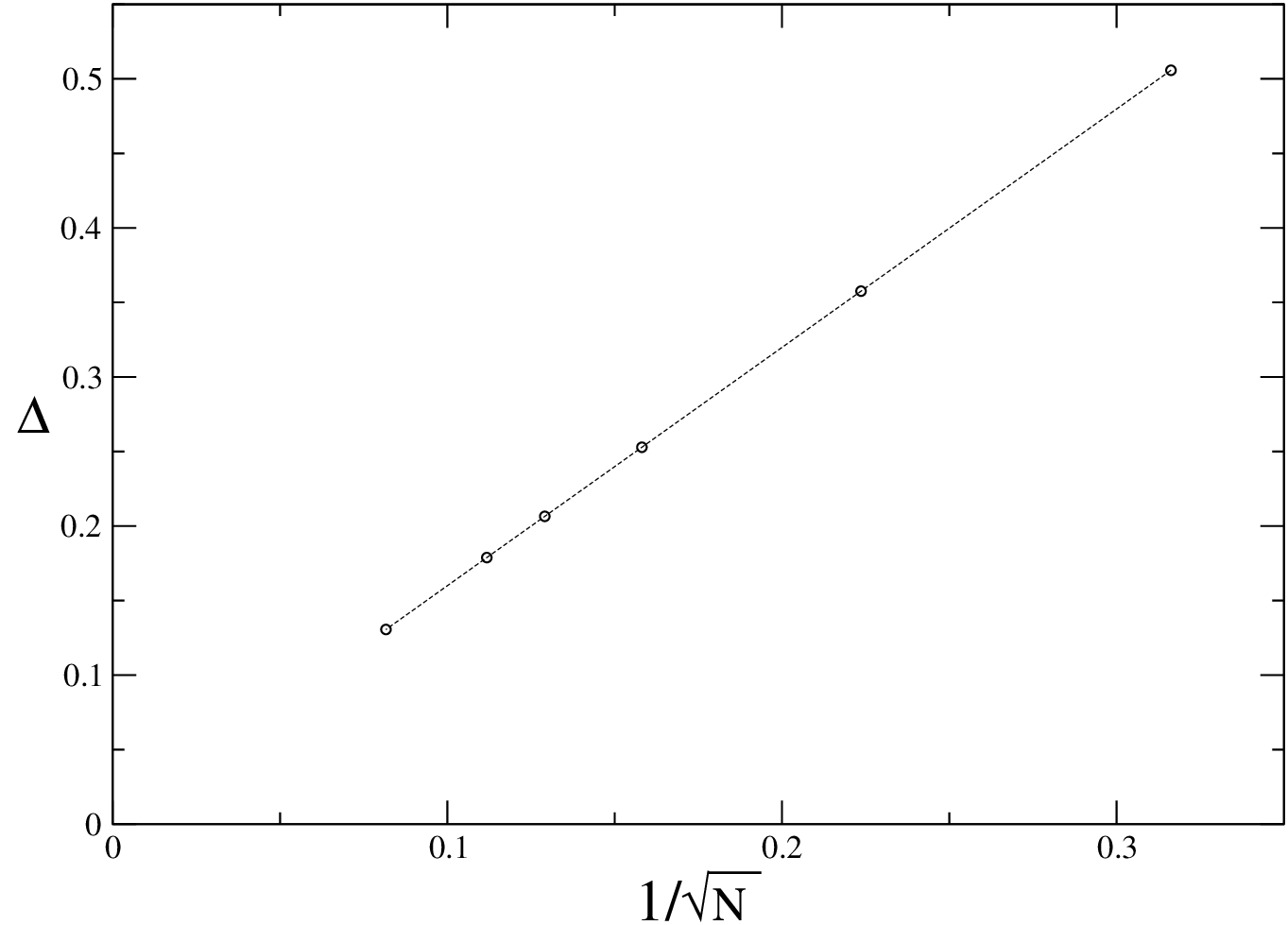}
\end{center}
\caption{Resonance width as a function of $1/\sqrt{N}$. The points are results of the
numerical calculation. The dashed line corresponds to the linear regression of the data.}
\label{hyperbola}
\end{figure} \\
The dependence of the resonance width with $N$ is presented in
Fig. \ref{hyperbola}. The numerical calculation fits with a hyperbola defined
by $\Delta\sqrt{N}\simeq1.599$. This result is in agreement with
eq.(\ref{senoc}) when $P_{s}\left(  N,\delta\right)  \simeq\frac{1}{2}$ with
$\delta=\Delta$.

\section{Connection with a discrete time search algorithm}

\label{sec:discrete}
We now build a discrete version of the continuous time algorithm 
developed in the previous sections.
Note that the time $\tau$ can be expressed by $\tau=\pi/(2\Omega)=\sqrt{N}\omega_{sj}T/4$, 
where $T=2\pi/\omega_{sj}$ is the period of interaction potential, eq.(\ref{potential}).
Then one can use Floquet's theory \cite{Reichl} to obtain the matrix of the Floquet 
evolution operator $U_{F}=U_{F}(T)$.
If one defines an unitary evolution operator $U_{D}\equiv \left(U_{F}\right)^{\frac{\omega_{sj}}{4}}$,
the searched state is obtained by applying $U_{D}$ $O(\sqrt{N})$ times to the initial state. 
If $\omega_{sj}/4$ is not an integer number, we round it to the nearest
integer in the definition of $U_{D}$. \\
To construct the Floquet operator, we must look for the amplitudes of the
Floquet states that are also solutions of the eqs.(\ref{dynamics2}). 
These amplitudes have the form

\begin{equation}
a_{m}(t)=\exp\left(  -i(\lambda-\varepsilon_{m})t\right)  \left\langle
m\right|  \left.  \phi_{\lambda}(t)\right\rangle , \label{floquet}%
\end{equation}

where $\left|  \phi_{\lambda}(t)\right\rangle =$ $\left|  \phi_{\lambda
}(t+T)\right\rangle $ is the $\lambda^{th}$ Floquet state and $\lambda$ is the
corresponding Floquet parameter. As $\left|  \phi_{\lambda}(t)\right\rangle $
is periodic, it can be expanded as a Fourier series%

\begin{equation}
\left\langle m\right|  \left.  \phi_{\lambda}(t)\right\rangle =\sum
\limits_{k=-\infty}^{\infty}A_{k}(m,\lambda)\exp\left(  il\omega_{sj}t\right)
. \label{Fourier}%
\end{equation}

Using the eqs (\ref{floquet},\ref{Fourier}) in eq.(\ref{dynamics2}) we obtain
set of equations for the parameter $\lambda$ and for the amplitudes $A_{k}$%

\begin{align}
\left(  -\lambda+\varepsilon_{j}+k\omega_{sj}\right)  A_{k}(j,\lambda)  &
=-\frac{1}{\sqrt{N}}\sum\limits_{m=1}^{N}A_{k+1}(m,\lambda)\text{,}%
\label{amplitud1}\\
\left(  -\lambda+\varepsilon_{n}+k\omega_{sj}\right)  A_{k}(n,\lambda)  &
=-\frac{1}{\sqrt{N}}A_{k-1}(j,\lambda)\text{, for all }n\in\mathbf{N}\text{.}
\label{amplitud2}%
\end{align}
Replacing $A_{k+1}$ from eq. (\ref{amplitud2}) into eq. (\ref{amplitud1}), the
characteristic equation for $\lambda$ is obtained%

\begin{equation}
xN=\sum\limits_{m=1}^{N}\frac{1}{x+\omega_{sm}}\text{,} \label{caracteristica}%
\end{equation}
where $x\equiv-\lambda+\varepsilon_{j}+k\omega_{sj}$. This equation has $N+1$
solutions for both $x$ and $\lambda$. Note that the indetermination in $\lambda$ introduced by
the term $k\omega_{sj}$ is superfluous since the Floquet evolution operator is evaluated 
at $t=T$ and therefore $\exp(-ik\omega_{sj}T)=1$. Once we have the values for $\lambda$ we
obtain the Fourier coefficients $A_{k}(m,\lambda)$ and, using eq.(\ref{Fourier}), 
$\left\langle m\right|  \left.  \phi_{\lambda}(t)\right\rangle $. \\

The Floquet matrix has the form
\begin{equation}
\left(U_{F}\right)_{mn}\left(  T\right)  =\sum\limits_{\lambda}\exp(-i\lambda
T)\left\langle n\right|  \left.  \phi_{\lambda}(0)\right\rangle \left\langle
\phi_{\lambda}(0)\right|  \left.  m\right\rangle \text{,}\label{evolu1}%
\end{equation}
and, therefore,
\begin{equation}
\left(U_{D}\right)_{mn}\left(  T_{0}\right)  =\sum\limits_{\lambda}\exp(-i\lambda
T_{0})\left\langle n\right|  \left.  \phi_{\lambda}(0)\right\rangle
\left\langle \phi_{\lambda}(0)\right|  \left.  m\right\rangle \text{,}%
\label{evolu2}%
\end{equation}
where $T_{0}=\omega_{sj}T/4$. Interpreting  $U_{D}$ as
a rotation operation with angle $\lambda T_{0}$, its application to the
initial state $O(\sqrt{N})$ times, maximizes the probability to find the searched state. 
Then for any $H_{0}$ one can build a discrete time algorithm to perform the search.
 
\section{Conclusions}

\label{sec:conclusion} We have developed a new insight in generating a
continuous time quantum search algorithm using a characteristic of quantum
mechanics, quantum resonances. This algorithm behaves like Grover's algorithm; 
in particular the optimal search time is proportional to
the square root of the size of the search set, $\sqrt{N}$, and the probability
to find the searched state oscillates periodically in time. The efficiency of
this algorithm depends on the spectral density of the Hamiltonian $H_{0}$. A
bigger separation between the energy levels maximizes the probability of the
searched state and allows for a better precision. For any Hamiltonian with
discrete energy spectrum, independently of its spectral density, this
algorithm can be implemented taking a large enough search set. 

The algorithm was shown to be robust when the energy of the searched state has
some imprecision. However the improvement in the accuracy of the search by
increasing $N$, and the error margin are bounded by the relation $\Delta
\sqrt{N}=$ constant. This means that for a large $N$ it demands a good
precision in the eigenvalue of the searched state. Noting that $\Delta$ is an
energy variation, and $\sqrt{N}$ is the time needed for the measurement, the
previous relation is simply the Heisenberg uncertainty principle. \\ 
\newline
We have found a simple way to build a discrete time algorithm on the basis of our
continuous time search algorithm, which strongly suggests the equivalence between
search algorithms with discrete and continuous time. \\
\newline 
It has been recently shown that Grover's algorithm may not be directly applicable 
to search in a physical database\cite{Childs}. Indeed, it has also been pointed out 
that it would be neither technologically nor economically reasonable to build 
database search engines basen on Grover's search algorithm\cite{Zalka}. 
Similar questions might be raised about the algorithms presented here.

Regarding the implementation of these algorithms, we note that the quantum kicked 
rotor has been experimentally realized using ultra-cold atom traps and
some experiments have focused on the resonant case \cite{williams}. Furthermore,
we have recently shown that the discrete quantum random walk on the
line has the same dynamics as the kicked rotor in resonance
\cite{Hadamard,hadarotor}. Several systems have been proposed as candidates to
implement quantum random walks \cite{Dur,Sanders,Du,Berman,Roldan}. When these
devices are constructed, they may be employed to simulate the search algorithms
proposed here.

We acknowledge the comments made by V.  Micenmacher and the support from PEDECIBA
and PDT S/C/OP/28/84. R.D.  acknowledges partial financial support from the
Brazilian National Research Council (CNPq) and FAPERJ (Brazil).
A.R and R.D. acknowledge financial support from the \textit{Brazilian Millennium
Institute for Quantum Information}.


\begin{thebibliography}{99}
\bibitem{Feynman}R. Feynman, \textit{Int. J. Theor. Phys.} \textbf{21}, 467 (1982).

\bibitem{Chuang}M. Nielssen and I. Chuang, \textit{Quantum Computation and
Quantum Information}, Cambridge University Press, 2000.

\bibitem{Shor}P.W. Shor, Proc. of the 35$^{th}$ Annual Symposium on the
Foundations of Computer Science, Ed. S. Goldwasser, Los Alamitos, CA, 1994;
\textit{ibid.} SIAM J. Comp., \textbf{26}, 1484, (1997).

\bibitem{Grover}L. K. Grover, Proc. 28$^{th}$ STOC, 212, Philadelphia, PA
(1996) and L.K. Grover, Phys. Rev. Lett. \textbf{79}, 325 (1997)

\bibitem{Boyer} M. Boyer, G. Brassard, P. H{\o}yer, and A. Tapp,
\textit{Fortsch. Phys.} \textbf{46} (1998) 493, also in quant-ph/9605034.

\bibitem{Farhi}E. Farhi and S. Gutmann. \textit{Phys. Rev}. A \textbf{57},
2403 (1998)

\bibitem{Childs}A. M. Childs and J. Goldstone \textit{Phys. Rev}. A
\textbf{70}, 022314 (2004) also in quant-ph/0306054, quant-ph/0405120

\bibitem{Grover2}L. K. Grover, A.M. Sengupta, \textit{Phys. Rev}. A \textbf{64},
032319 (2002), quant-ph/0109123.

\bibitem{Zak}M. Zak, in \textit{Quantum Computing and Quantum Communications},
(Lecture Notes in Computer Science 1509), Springer-Verlag,1999

\bibitem{Reichl}L. E: Reichl, \textit{The Transition to Chaos, In Conservative
Classical Systems: Quantum Manifestations, }Springer-Verlag,1992

\bibitem{Zalka}C.Zalka, \textit{Phys. Rev. A} \textbf{62}, 052305, (2000).

\bibitem{williams}M.E.K. Williams, M.P. Sadgrove, A.J. Daley, R.N.C. Gray,
S.M. Tan, A.S. Parkins, R. Leonhardt, quant-ph/0209090 (2002); F.L. Moore,
J.C. Robinson, C.F. Bharucha, B. Sundaram, M.G. Raizen, \textit{Phys. Rev.
Lett.} \textbf{75}, 4598 (1995).

\bibitem{Hadamard}A. Romanelli, A.C. Sicardi-Schifino, R. Siri, G. Abal, A.
Auyuanet, R. Donangelo. \textit{Physica.} A, \textbf{338}, 395 (2004) also in quant-ph/0310171.

\bibitem{hadarotor}A. Romanelli, A. Auyuanet, R. Siri, G. Abal, R. Donangelo.
\textit{Physica.} A, \textbf{347}, 137 (2005) also in quant-ph/0408183.

\bibitem{Dur}W. D\"{u}r, R. Raussendorf, V. Kendon and H. Briegel, Phys. Rev.
A \textbf{66}, 052319 (2002); arXiv:quant-ph/0207137.

\bibitem{Sanders}B. Sanders, S. Bartlett, B. Tregenna and P. Knight, Phys.
Rev. A \textbf{67}, 042305 (2003); arXiv:quant-ph/0207028.

\bibitem{Du}J. Du, X. Xu, M. Shi, J. Wu, X. Zhou and R. Han, Phys. Rev. A
\textbf{67}, 042316 (2003).

\bibitem{Berman}G.P. Berman, D.I. Kamenev, R.B. Kassman, C. Pineda and V.I.
Tsifrinovich, Int. J. Quant. Inf. \textbf{1}, 51 (2003); arXiv:quant-ph/0212070

\bibitem{Roldan}P.L. Knight, E. Rold\'{a}n and E. Sipe, Phys. Rev. A
\textbf{68} 020301(R) (2003); \textit{ibid.} Optics Comm. \textbf{227}, 147 (2003).
\end{thebibliography}
\end{document}